\documentstyle[11pt,hrd_pasp,epsfig]{article}
\def\edcomment#1{\iffalse\marginpar{\raggedright\sl#1\/}\else\relax\fi}
\reversemarginpar

\begin{document}
\title{Tracing out the northern stream of the Sagittarius dwarf galaxy with color-magnitude diagram techniques}
 \author{D. Mart\'\i nez-Delgado}
\affil{Instituto de Astrof\'\i sica de Canarias, La Laguna, Tenerife, Spain}
\author{M. A. G\'omez-Flechoso}
\affil{Geneva Observatory, CH-1290 Sauverny, Switzerland}

\author{A. Aparicio}
\affil{Instituto de Astrof\'\i sica de Canarias, La Laguna, Tenerife, Spain}

\begin{abstract}

Standard cosmology predicts that dwarfs were the first galaxies to be formed
in the Universe and that many of them merge afterwards to form bigger galaxies
such as the Milky Way. This process would have left behind traces such as tidal
debris or star streams in the outer halo. We report here the detection of two
new tidal debris of the northern stream of the Sagittarius dwarf galaxy, based in the analysis of wide field, deep color-magnitude diagrams. These detections provide strong observational evidence  that the stripped debris of Sagittarius extends up to $60\deg$ from its center, suggesting that the stream of this galaxy completely wraps the Milky Way in an almost polar orbit.  Our negative detections also suggest the stream is narrow, supporting a nearly spherical Milky Way dark matter halo potential.

\end{abstract}

\section{Introduction}

A result of standard cosmology is that dwarf protogalaxies are the first to
born as individual systems in the Universe. Afterwards, many of these merge to
form larger galaxies such as the Milky Way. The way in which this process
takes place has consequences for the present-day structure of the Milky
Way. The significant issues (White \& Frenk 1991) are how the merging
efficiency compares with the star formation efficiency in the protogalactic
fragments, and how the fragment merging and disruption time-scales compare
with the age of the Milky Way.  

Of particular relevance has been the discovery of the Sagittarius (Sgr) dwarf galaxy (Ibata, Gilmore, \& Irwin 1994), a Milky Way satellite in an advanced state of tidal disruption which provides a ``living'' test for tidal
interaction models and for galaxy formation theories. It was soon apparent that its extent was larger than at first assumed, and dynamical models predict that the stream associated
with the galaxy should envelop the whole Milky Way in an almost polar orbit.

Recently, two teams of the Sloan Digital Sky Survey (SDSS), have reported the detection of two clear $\sim45\deg$ long parallel strips of blue
A-type stars, that are the trace of a stream in the outer Galactic halo
(Yanny et al. 2000). Mart\'\i nez-
Delgado et al. (2001) presented the first color--magnitude diagram (CMD) of this structure, finding
the evidence of a very low density stellar system at 50 $\pm$ 10 kpc from the Galactic center
and at 60$\deg$ away from the center of Sgr galaxy.
Theoretical models of Sgr (Ibata et al. 2001a; Mart\'\i nez-Delgado et al. 2001) have identified this structure as part of the northern stream of the Sgr dwarf galaxy. In this paper we report the first
detection of Sgr stellar members between the aforementioned SDSS stream and the Sgr main body.

\begin{figure}  
\plotone{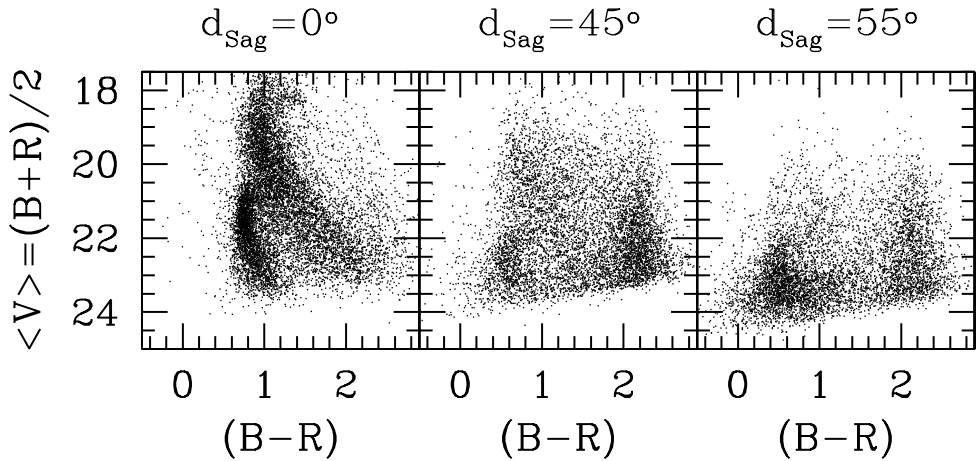}
\caption{ CMDs for the the central region of Sgr (d=$0\deg$) and two
regions of its northern stream situated at angular distances of $45\deg$ and $55\deg$ from its center.}
\end{figure}

\section{Methodology}

The detection of Sgr northern stream is very challenging due to
its large angular size and low surface brightness (LSB), and requires
using wide-field observations and a careful analysis of the foreground and
background contamination. An efficient technique is based on the analysis of
wide-field, deep CMDs (e.g., Mateo,
Olszewski, \& Morrison 1998: Mart\'\i nez-Delgado et al. 2001). The LSB tail can be detected through star counts
at the old population main-sequence (MS) turn-off region. This feature is the
most densely populated in the CMD and thus provides the best contrast against
the foreground and background population.

The observations of Sgr tidal stream were carried out in $B$ Johnson--Cousins
filter and $R$ Sloan filter with the Wide Field Camera at 2.5 m Isaac Newton Telescope (INT) at Roque
de los Muchachos Observatory on La Palma. For the Sgr northern stream position is poorly defined in the surveyed area, a very important aspect of this work is to alter the search pattern during the observation. This means doing 'real-time' photometry as soon as the data
are obtained (see Mateo et al. 1998). This procedure was performed using DAOPHOT/ALLSTAR software in the Ultra workstation at the telescope.

\begin{figure}  
\plotone{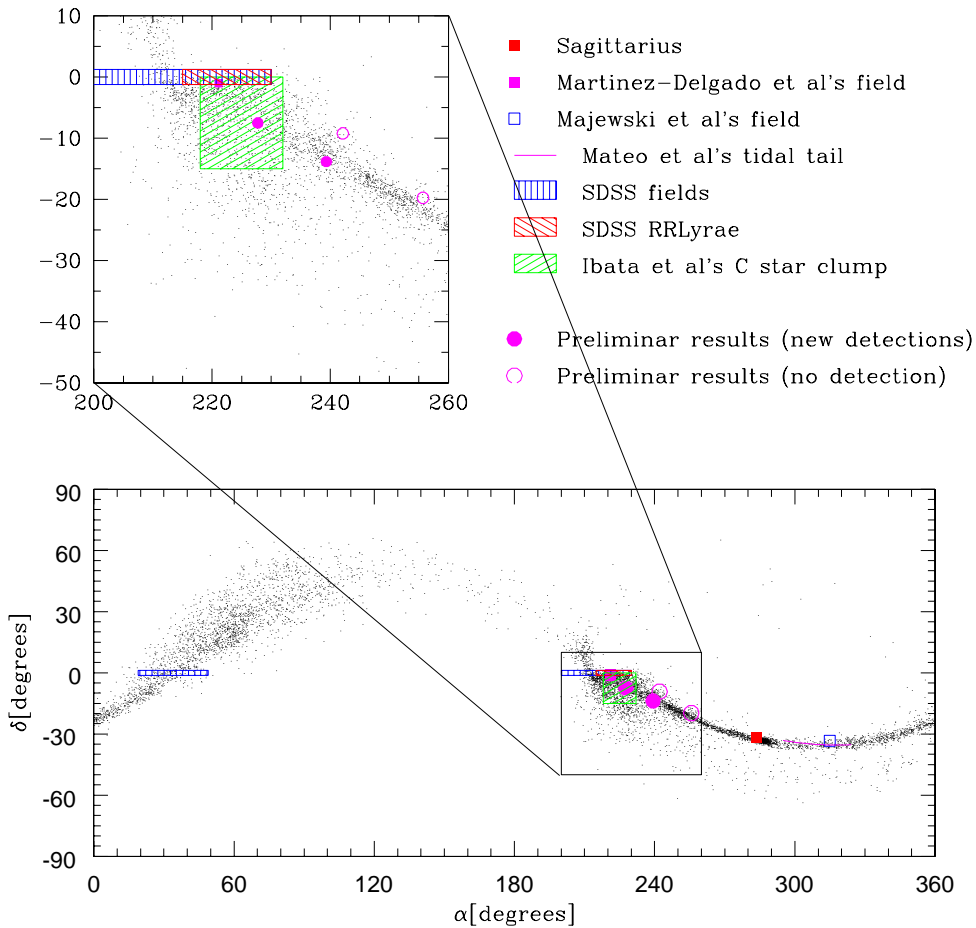}
\caption{ Equatorial coordinates of our model for Sgr and the observational detections of
the Sgr northern stream. }
\end{figure}

\section{Results}

The main result of our survey is the detection of two new tidal debris of the Sgr northern stream, outside the region covered by the SDSS. Figure 1 shows the CMDs for the center of the galaxy and for these two new tidal debris discovered along the Sgr northern stream. The
magnitude of the MS turnoff of Sgr increases with the distance to its center, confirming that the the distance of the stream to the Milky Way center increases as we move away from Sgr center. The derived distances for these tidal debris at 45$\deg$ and 55$\deg$ are 32 kpc and 45 kpc respectively, in good agreement with our theoretical model (Mart\'\i nez-Delgado et al.
2001). The surface brightness of the stream increases at the galaxy's apocenter, as it is also expected from the model predictions. These detections are strong observational evidence 
that the SDSS stream is tidally stripped material from Sgr dwarf and support the idea that this galaxy completely wraps the Milky Way in an almost polar orbit.

	The tidal debris at 45$\deg$ overlaps with the prominent clump of carbon stars discovered by  Ibata et al. (2001). The kinematic of these stars is also consistent with
the radial velocities of stream's red giants by Dohm-Palmer et al.(2001). These results
confirm that these carbon stars are tracing the intermediate-population of the densest part of the stream. In turn, this suggests that the debris stellar
population shares the complex star formation history observed in the center region of Sgr (Mart\'\i nez-Delgado et al. 2002). To understand the properties of the stellar population at this region is fundamental to constrain the origin and evolution of the Sgr stream. 

	 We also reported two negative detections of the stream at lower
galactic latitudes. This indicates that the Sgr stream is shifted by several
degrees in galactic longitude ($l$) from its expected position at this
galactic latitude. This result will be very useful to better constrain the
orbit of Sgr and indicates that the inclination of the stream plane is not
as close to 90$\deg$ (with respect to the Milky Way plane) as previously 
thought.

	Figure 2 shows the projection on the sky of our update Sgr tidal tail model (see Mart\'\i nez-Delgado et al. 2001). Our new positive and negative detections are indicated together with the previous observational data of the northern stream.  The negative detections suggest a narrow stream, supporting a nearly spherical Milky
Way dark matter halo potential. From our simulations, we obtain an axial radial of the Milky Way dark halo potential of $q_{DH}=0.8$ (see Mart\'\i nez-Delgado et al. 2002).  We are carrying out a new survey to constrain the width of the Sgr stream at low galactic latitudes and improve our determination of this parameter.

\end{document}